\documentclass[rmp,showpacs]{revtex4}
\usepackage{amsmath,amssymb}
\begin{document}
\def\cs#1#2{#1_{\!{}_#2}}
\def\css#1#2#3{#1^{#2}_{\!{}_#3}}
\def\ocite#1{[\onlinecite{#1}]}
\def\ket#1{|#1\rangle}
\def\bra#1{\langle#1|}
\def\expac#1{\langle#1\rangle}
\def\dbl{\hbox{${1\hskip -2.4pt{\rm l}}$}}
\def\bfh#1{\bf{\hat#1}}

\title{Disproof of Bell's Theorem: Further Consolidations}

\author{Joy Christian}

\email{joy.christian@wolfson.oxford.ac.uk}

\affiliation{Perimeter Institute, 31 Caroline Street North, Waterloo, Ontario N2L 2Y5, Canada,}
\affiliation{Department of Physics, University of Oxford, Parks Road, Oxford OX1 3PU, England}


\begin{abstract}
The failure of Bell's theorem for Clifford algebra valued local variables is further
consolidated by proving that the conditions of remote parameter independence and remote
outcome independence are duly respected within the recently constructed exact, local realistic
model for the EPR-Bohm correlations. Since the conjunction of these two conditions is
equivalent to the locality condition of Bell, this provides an independent geometric proof 
of the local causality of the model, at the level of microstates. In addition to local
causality, the model respects at least seven other conceptual and operational requirements,
arising either from the predictions of quantum mechanics or the premises of Bell's theorem,
including the Malus's law for sequential spin measurements. Since the agreement between the
predictions of the model and those of quantum mechanics is quantitatively precise in all
respects, the ensemble interpretation of the entangled singlet state becomes amenable.

\end{abstract}

\maketitle

\parskip 5pt

\section{Introduction}

One of the sources of the geometrical beauty, internal coherence, and empirical success of
general relativity is its strict adherence to local causality \ocite{local}. Quantum theory,
on the other hand, is peculiarly defiant of this notion, as was stressed many years ago
by Einstein, Podolsky, and Rosen (EPR) \ocite{EPR}. Worse still, any hope of ``completing''
quantum theory into a realistic, locally causal theory in a manner espoused by EPR is
generally believed to have long been dashed by Bell's theorem and its variants
\ocite{Bell-1964}\ocite{GHSZ}. Indeed, these theorems are breathtakingly ambitious in their
scope to frustrate any attempt to the contrary: {\it no physical theory which is realistic
as well as local in a specified sense can reproduce all of the statistical predictions of 
quantum mechanics} \ocite{Shimony-2004}.
In a recent paper \ocite{Christian}, however, the
legitimacy of this claim was put in doubt by means of an exact, deterministic, local
realistic model for the EPR-Bohm correlations on which Bell's theorem rests, without
appealing to either remote contextuality or backward causation. In particular, it was shown
that the much studied CHSH inequality \ocite{Clauser} in this context is violated within this
model and extended to the extrema of ${\pm\,2\sqrt{2}\,}$, in {\it exactly} the same manner
as it is within quantum mechanics. Despite these compelling features, however, since it
stands against the received wisdom of some four decades, the model has met with a certain
predisposed skepticism. In what follows this skepticism is systematically addressed, and proven
to be entirely unwarranted. This is accomplished by demonstrating that, not only the model
adheres {\it strictly} to the notion of local causality as prescribed by Bell, but also the
agreement between the predictions of the model and those of quantum mechanics is
{\it quantitatively} precise in {\it all} relevant aspects of physics, within the context
of his theorem.

\subsection{Exact, Deterministic, Locally Causal Model for the EPR-Bohm Correlations}

The central strategy of the proposed model of Ref.\ocite{Christian} is to make use of the
Clifford algebra valued observables in a manner that renders the non-commuting parts of their 
products to contribute only counterfactually. The complete state describing the singlet spin 
system is taken to be a unit trivector ${\boldsymbol\mu}$,
which is an element of the Clifford algebra ${Cl_{3,0}}$ of the subspaces of the Euclidean space
${{\rm I\!R}^3}$. It can be viewed as a unit volume element in ${{\rm I\!R}^3}$, assembled by
three ordinary vectors of finite lengths and arbitrary directions. Since every trivector in the
algebra ${Cl_{3,0}}$ differs only by its volume and orientation from the standard trivector
${I}$ composed of a right-handed frame of orthonormal vectors, ${\boldsymbol\mu}$ differs from
${I}$ only by the sign of its handedness: ${{\boldsymbol\mu}=\pm\,I}$. Thus, the local hidden
variable of the model is essentially the intrinsic freedom of choice in the initial orientation
of the volume element ${\boldsymbol\mu}$, which is a pseudoscalar, dual to a scalar in
${{\rm I\!R}^3}$, and hence commutes with every other element of ${Cl_{3,0}}$.  Next, the actual
spin observables in the model are taken to be the projections
${{\boldsymbol\mu}\cdot{\bf n}=:A_{\bf n}(\boldsymbol\mu)}$ of the trivector ${\boldsymbol\mu}$
along the unit directions ${\bf n}$, which are simply the directions of the  orientations of
spin analyzers. In other words, the observables ${A_{\bf n}(\boldsymbol\mu)}$ are the unit,
non-commuting bivectors ${{\boldsymbol\mu}\cdot{\bf n}}$, isomorphic to the familiar
quaternionic numbers \ocite{Doran}. Finally, the Clifford product of two such observables, say 
${A_{\bf a}(\boldsymbol\mu)}$ and ${A_{\bf b}(\boldsymbol\mu)}$, satisfies the following
identity, which plays a central role in the model:
\begin{equation}
(\,{\boldsymbol\mu}\cdot{\bf a})(\,{\boldsymbol\mu}\cdot{\bf b})\,
=\,-\,{\bf a}\cdot{\bf b}\,-\,{\boldsymbol\mu}\cdot({\bf a}\times{\bf b}).\label{bi-identity}
\end{equation}
This identity is simply a more general expression of the bivector subalgebra within the algebra
${Cl_{3,0}}$ (cf. Eq.${\,}$(2.65) of Ref.\ocite{Doran}). For notations, conventions, and further
background, the reader is urged to consult Refs.\ocite{Christian} and \ocite{reply}.

\subsection{Eight Essential Requirements Satisfied by the Local Model}

As elementary as this model seems to be, it respects at least {\it eight} different conceptual
and operational requirements, arising from either the predictions of quantum mechanics or the
premises of Bell's theorem. For convenience, let us begin our analysis by anthologizing these
requirements, without any attempt to iron out redundancies:

{\bf (1)} Mathematical representation of the physical quantities in a proposed local model for
the EPR-Bohm correlations must not be {\it ad hoc} \ocite{Bell-1971}, but a part of at least
operationally well-motivated theoretical framework. It will be clear from the discussion
in the next section that our choice of representation for the observables
${A_{\bf n}(\boldsymbol\mu)}$ as elements of the Clifford algebra ${Cl_{3,0}}$ amply satisfies 
this requirement. Since the pioneering works of Grassmann and Clifford, there has been a
sustained impetus to reformulate {\it all} physical quantities within the Clifford algebraic
framework \ocite{Oersted}\ocite{Clifford}\ocite{Dorst}. What is more, it is known that
many of the quantum observables find natural expressions within this framework
\ocite{Doran}\ocite{Oersted}. The physical motivations behind our own choice of the local
realistic observables within this framework has been made clear in Ref.\ocite{Christian} (see
also Ref.\ocite{reply}).
Below we will discuss why this choice is also operationally well motivated. 

{\bf (2)} The actual values of the observables ${A_{\bf n}(\boldsymbol\mu)}$, for all
${\bf n}$ and ${\boldsymbol\mu}$, must lie in the interval ${[-1,\,+1]}$. From the discussion
below, it will be clear that this requirement is necessarily satisfied by the observables of
our model, despite the non-commutativity of their products. The only non-zero values the
observables ${A_{\bf n}(\boldsymbol\mu)}$ can possibly yield in any experiment correspond
to a {\it bit} of information. Thus ${A_{\bf n}(\boldsymbol\mu)}$ are truly dichotomic
observables, with binary values ${\pm\,1}$.

{\bf (3)} The classical ensemble average of a single observable
${A_{\bf n}(\boldsymbol\mu)}$, taken as a dispersion-free counterpart \ocite{Bell-1966}
of the corresponding quantum mechanical spin operator ${{\boldsymbol\sigma}\cdot{\bf n}}$,
must be equal to the quantum mechanical expectation value of the operator
${{\boldsymbol\sigma}\cdot{\bf n}\,}$ in the singlet spin state ${\ket{\Psi_{\bf n}}}$.
In the notations of Ref.\ocite{Christian} this statement can be expressed as
\begin{equation}
{\cal E}_{h.v.}({\bf n})
\;=\int_{{\cal V}_3}A_{\bf n}(\boldsymbol\mu)\;\,d{\boldsymbol\rho}(\boldsymbol\mu)\,=\,
\bra{\Psi_{\bf n}}\,{\boldsymbol\sigma}\cdot{\bf n}\,\ket{\Psi_{\bf n}}\,=\,0\,.\label{zero}
\end{equation}
It is demonstrated in Ref.\ocite{Christian} that this result---which turns out to be a
straightforward consequence of the dichotomic nature of the observables 
${A_{\bf n}(\boldsymbol\mu)}$---holds exactly within our local model.

{\bf (4)} The classical ensemble average of a joint observable
${(A_{\bf a}\,B_{\bf b})(\boldsymbol\mu)}$, satisfying the factorizability condition
\begin{equation}
(A_{\bf a}\,B_{\bf b})(\boldsymbol\mu)\,=\,
A_{\bf a}(\boldsymbol\mu)\,B_{\bf b}(\boldsymbol\mu)\,,\label{factori}
\end{equation}
and taken as a dispersion-free counterpart of the corresponding quantum mechanical spin
operator ${{\boldsymbol\sigma}_1\cdot{\bf a}\,\otimes\,{\boldsymbol\sigma}_2\cdot{\bf b}}$,
must be equal to the quantum mechanical expectation value of this operator
in the singlet state ${\ket{\Psi_{\bf n}}}$, giving
\begin{equation}
{\cal E}_{h.v.}({\bf a},\,{\bf b})
\;=\int_{{\cal V}_3}A_{\bf a}(\boldsymbol\mu)\;B_{\bf b}(\boldsymbol\mu)
\;\,d{\boldsymbol\rho}({\boldsymbol\mu})\,=\,
\langle\Psi_{\bf n}|\,{\boldsymbol\sigma}_1\cdot{\bf a}\,\otimes\,
{\boldsymbol\sigma}_2\cdot{\bf b}\,|\Psi_{\bf n}\rangle\,=\,-\,
{\bf a}\cdot{\bf b}\,.\label{twoobserve}
\end{equation}
As demonstrated in Ref.\ocite{Christian}, this result---which follows almost trivially from the
identity (\ref{bi-identity})---holds {\it exactly} within our local model. It is also worth
stressing here that the condition ${{\cal E}_{h.v.}({\bf n},\,{\bf n}) = -\,1\,}$, which was
adapted by EPR for the special and ideal case of perfect correlation, and which also played a
crucial role in the original theorem of Bell concerning only deterministic local hidden
variable theories \ocite{Bell-1964}, also holds exactly within our local model.

{\bf (5)} In conjunction with the factorizability condition (\ref{factori}) above, the actual
values  of the observables ${A_{\bf a}(\boldsymbol\mu)}$, ${B_{\bf b}(\boldsymbol\mu)}$,
and ${(A_{\bf a}\,B_{\bf b})(\boldsymbol\mu)}$---for all ${\bf a}$, ${\bf b}$, and
${\boldsymbol\mu}$---must respect the following relationship among themselves:
\begin{equation}
v(A,\,B)\,=\,A\,B\,,\label{loc-00}
\end{equation}
where ${A}$ and ${B}$, respectively, are the actual values that could be found if the observables
${A_{\bf a}(\boldsymbol\mu)}$ and ${B_{\bf b}(\boldsymbol\mu)}$ are measured in some experiment,
and ${v(A,\,B)}$ is the corresponding value of the joint observable
${(A_{\bf a}\,B_{\bf b})(\boldsymbol\mu)}$. In the subsection III below, this condition---which
is simply the celebrated locality condition of Bell
\ocite{Bell-1964}\ocite{Clauser-Shimony}---will be shown to hold rigorously within our local
model. It can be restated in two conceptually distinct parts as follows:

{\bf (5${\,}$a)} A proposed local model must satisfy the condition of remote parameter
independence \ocite{Shimony-2004}. The deterministic version of this part of the locality
condition states that, for a given microstate ${\boldsymbol\mu}$, the outcome of an experiment
at a local station {\bf 1} must be independent of the chosen experimental settings at the remote
station {\bf 2}, and vice versa. In the subsection III A below, this condition will be shown to
hold rigorously within our local model.

{\bf (5${\,}$b)} A proposed local model must satisfy the condition of remote outcome independence
\ocite{Shimony-2004}. The deterministic version of this part of the locality condition
states that, for a given microstate ${\boldsymbol\mu}$, the outcome of an experiment at a local
station {\bf 1} must be independent of the outcome of an experiment at the remote station
{\bf 2}, and vice versa. In the subsection
III B below, this condition will be shown to hold rigorously within our local model.

{\bf (6)} The choice of the measurement settings within a proposed local model for the
EPR-Bohm correlations, such as our vectors ${\bf a}$ and ${\bf b}$, should not in any way be
constrained by the microstates ${\boldsymbol\mu}$. More pertinently, the distribution
${{\boldsymbol\rho}(\boldsymbol\mu)}$ of the microstates must be independent of the
measurement settings
${\bf a}$ and ${\bf b}$ \ocite{Bell-La}. As discussed in Ref.\ocite{Christian}, this requirement,
accommodating the ``free will'' of the experimenter, is well respected within our local model.

{\bf (7)} A proposed model for the EPR-Bohm correlations must violate the
CHSH inequality in precisely the same quantitative manner as do the corresponding nonlocal
predictions of quantum mechanics. Namely, in the standard notations \ocite{Clauser-Shimony},
the correlation functions (\ref{twoobserve}) must satisfy the
Tsirel'son inequality \ocite{bounds}:
\begin{equation}
\left|\,{\cal E}({\bf a},\,{\bf b})\,+\,{\cal E}({\bf a},\,{\bf b'})\,+\,{\cal E}({\bf a'},\,
{\bf b})\,-\,{\cal E}({\bf a'},\,{\bf b'})\,\right|\,\leqslant\,2\sqrt{2}\,.\label{My-CHSH}
\end{equation}
That this inequality is exactly predicted by our local model has been already exhibited
twice. In Ref.\ocite{Christian} it was shown to hold in general terms within the model. This
demonstration, however, proved to be too abstract for some readers, and hence, in response
\ocite{reply}, the inequality was again derived in terms of the cosines of relative angles
between the four spin analyzers. In section IV below, we shall once again derive this inequality
in a different manner, to exhibit the model's precise quantitative agreement with the
corresponding predictions based on the entangled singlet state.

{\bf (8)} A complete local model for the EPR-Bohm correlations must also reproduce the Malus's
law for sequential spin measurements, since quantum mechanics respects this law. In our
notations, this requirement can be stated as
\begin{equation}
\int_{{\cal V}_3} A({\bf a},\,{\bf p},\,{\boldsymbol\mu})
\;\;d{\boldsymbol\rho}(\boldsymbol\mu)\,=\,
\bra{\,s_{\bf p}=\pm1\,}\,{\boldsymbol\sigma}\cdot{\bf a}\,\ket{\,s_{\bf p}=\pm1\,}\,
=\,{\bf a}\cdot{\bf p}\,,
\end{equation}
where the observable ${A({\bf a},\,{\bf p},\,{\boldsymbol\mu})}$ now depends on both the
polarizer ${\bf p}$ and the analyzer ${\bf a}$. In section V below, this relation, as well as
its extension to sequential spin measurements, will be shown to hold exactly within our model.

The above results complement and strengthen the disproof of Bell's theorem presented in
Ref.\ocite{Christian}. More precisely, if we take Bell's theorem to be the statement that:
{\it it is impossible to construct a deterministic local hidden variable theory based on
the observables ${A_{\bf n}(\boldsymbol\mu)}$ and probability distribution
${{\boldsymbol\rho}(\boldsymbol\mu)}$ such that all eight of the requirements listed above are
satisfied in conjunction}, then the local realistic model constructed in Ref.\ocite{Christian}
decisively disproves this theorem.

To better appreciate this claim, let us first have a closer look at the basic observables of
our model.

\section{Operational adequacy of the observables ${A_{\bf n}(\boldsymbol\mu)}$, and a Theory
of their Measurement:}

As we noted above, the local observables ${A_{\bf a}(\boldsymbol\mu)}$ and
${B_{\bf b}(\boldsymbol\mu)}$ of our model are unit bivectors, with binary values ${\pm\,1}$.
Some readers and commentators have found the use of non-commuting bivectors as observables
less than compelling \ocite{reply}.
In the proof of Bell's theorem, they allege, locality requires that the observables such as
${A_{\bf a}(\boldsymbol\mu)}$ and ${B_{\bf b}(\boldsymbol\mu)}$ must simply be ordinary
functions---maps, if you like, from whatever valued hidden variables ${\boldsymbol\mu}$
one may wish to consider, to commuting real numbers. Indeed, they argue, in the derivation
of CHSH inequality one only needs to assume that the outcomes ${A_{\bf a}(\boldsymbol\mu)}$
and ${B_{\bf b}(\boldsymbol\mu)}$---for all ${\bf a}$, ${\bf b}$, and ${\boldsymbol\mu}$---are 
commuting real numbers, lying in the closed interval ${[-1,\,+1]}$. In this section we
point out that the above charge against our local model is based on a failure to appreciate 
the Clifford algebraic concepts employed in the model, and that it overplays the necessity
of commuting numbers within the operational contexts of Bell's theorem \ocite{reply}. In the
following section we then demonstrate that the relativistic local causality---as famously
crystallized by Bell in his locality (or factorizability)
condition \ocite{Bell-La}---is rigorously
respected within our model, despite the apparent non-commutativity of our realistic
observables ${A_{\bf a}(\boldsymbol\mu)}$ and ${B_{\bf b}(\boldsymbol\mu)}$.

To this end, let us first recall that, since the pioneering vision of Grassmann circa 1844,
the very purpose of the Clifford algebra ${Cl_{3,0}}$ has been to treat {\it all} of the
subspace elements of ${{\rm I\!R}^3}$---namely, scalars, vectors, bivectors, and
trivectors---on equal footing, and use them as {\it direct} elements of geometric
computations \ocite{Doran}\ocite{Oersted}\ocite{Clifford}\ocite{Dorst}.
This provides a powerful computational framework that utilizes subspace elements across
dimensions, from 0-dimensional scalars to 3-dimensional trivectors, all equally respected
as ``directed numbers'', and unified by the {\it invertible} Clifford
product of immense power and versatility. In particular, within this framework scalars are
just as much a part of the algebra as vectors, bivectors, and trivectors are \ocite{Dorst}.
Indeed, a generic multivector in this 8-dimensional linear space ${Cl_{3,0}}$ includes a
scalar, and is written as a linear combination of that scalar, plus a vector, a bivector,
and a trivector:
\begin{equation}
{\boldsymbol\xi}\,=\,{\rm scalar}\,+\,{\rm vector}\,+\,{\rm bivector}\,+\,{\rm trivector}.
\end{equation} 
What is more, the familiar tools such as complex numbers and quaternions are now absorbed
within a {\it real} vector space, allowing one to treat them as {\it reals}, in every sense
of that word. Even the differential forms, symplectic forms, tensors, spinors, and matrices
are not spared from assimilation and simplification by this vastly versatile framework. Indeed,
the modern program of Clifford algebra is anything but modest. It aims to provide nothing less
than a {\it universal} language for all of mathematics, physics, and engineering
\ocite{Oersted}\ocite{Dorst}\ocite{Lasenby}. Unfortunately, as mentioned above, unfamiliarity
with this empowering and unifying framework has given rise to some misplaced questions and
unsubstantiated presumptions about our counterexample to Bell's theorem. In what follows, we
answer these questions by first spelling out exactly how do the observables
${A_{\bf n}(\boldsymbol\mu)}$ in our local model represent
what is actually observed in a Stern-Gerlach type experiment, and how do they do so more
faithfully than the observables of Bell's own local model.

As we noted above, the local realistic physical picture underlying our model is as
follows: By construction, the complete microstates are the {\it unit} volume elements
${\boldsymbol\mu}$, with unspecified orientations---i.e., the initial handedness of the
trivectors ${\boldsymbol\mu}$ is taken to be unspecified. Given two arbitrary unit vectors
${\bf a}$ and ${\bf b}$, specifying, say, the ``${z}$ directions'' of two Stern-Gerlach
apparatuses, the two remote spin observables for a given microstate are represented
by the bivectors ${A_{\bf a}(\boldsymbol\mu)={\boldsymbol\mu}\cdot{\bf a}}$ and
${B_{\bf b}(\boldsymbol\mu)={\boldsymbol\mu}\cdot{\bf b}}$, respectively. Now, since any
bivector has an intrinsic sense of rotation, spin is routinely represented in Clifford
algebra by a bivector (see, e.g., Eq.${\,}$(88) of Ref.\ocite{Oersted}). Moreover, 
the bivectors ${A_{\bf a}(\boldsymbol\mu)}$ and ${B_{\bf b}(\boldsymbol\mu)}$ of our model 
are necessarily of {\it unit} magnitude, which can be checked by evaluating their
norm: ${||A_{\bf n}(\boldsymbol\mu)||^2=||{\boldsymbol\mu}\cdot{\bf n}||^2=(\mp I\,{\bf n})
(\pm I\,{\bf n})={\bf n}\,{\bf n}={\bf n}\cdot{\bf n}=1}$. Thus, the magnitudes of
${A_{\bf a}(\boldsymbol\mu)}$ and ${B_{\bf b}(\boldsymbol\mu)}$ in our model are
universally fixed to be unity, and their directions---which are simply the normal
directions to their planes---are defined by the ``${z}$'' directions of
the Stern-Gerlach magnets themselves. This is because, just as the operator
${{\boldsymbol\sigma}\cdot{\bf n}}$ represents a projection of the spin operator
${\boldsymbol\sigma}$ along the direction ${\bf n}$ in quantum mechanics, the bivector
${{\boldsymbol\mu}\cdot{\bf n}}$ represents a projection of the volume element
${\boldsymbol\mu}$ along the direction ${\bf n}$ in our local model. And since geometrically
this projection gives rise to a bivector ${{\boldsymbol\mu}\cdot{\bf n}}$---which, in turn, is
simply a plane segment with an intrinsic sense of rotation perpendicular to the direction
${\bf n}$---the geometrical picture---reminiscent of a spinning wheel of a miniscule gyroscope
with its axis of rotation ${\bf n}$ aligned to the direction ${z}$ of the
Stern-Gerlach apparatus---can hardly be more compelling.

In fact, the local observables ${{\boldsymbol\mu}\cdot{\bf n}}$ are even operationally no less
compelling. To appreciate this, let us recall that, when considered by itself, each element
of Clifford algebra ${Cl_{3,0}}$ is {\it completely} specified by three properties, and three
properties {\it only} \ocite{Oersted}\ocite{Clifford}\ocite{Dorst}. These are (1) its
magnitude, (2) its direction, and (3) its orientation (or sense). In the case of our bivector
${{\boldsymbol\mu}\cdot{\bf n}}$, the first of these three properties is universally fixed,
once and for all. As we calculated above, its magnitude---which is simply the area of the plane
segment it represents---is universally fixed to be unity. And even the direction of this
bivector---which is specified by the direction of its dual vector ${\bf n}$---is fixed.
As discussed above, it is given by the direction of the spin analyzer itself,
once selected by the experimenter. Thus, once a choice of the direction ${\bf n}$ is made,
the direction of the bivector ${{\boldsymbol\mu}\cdot{\bf n}}$ is no longer unspecified.
Consequently, the only unspecified property of the bivector---i.e., the only unknown
property that a spin analyzer can possibly reveal---is its sense, inherited
from ${\boldsymbol\mu}$, which, for a bivector, is simply its sense of rotation---i.e., whether
it is spinning counterclockwise about ${\bf n}$ or clockwise---or, in the dual picture, whether
it is spinning ``up'' or ``down'' along ${\bf n}$. It is crucial to note here that a given
bivector ${{\boldsymbol\mu}\cdot{\bf n}}$ cannot be spinning either ``up'' or ``down'', or in 
any other way, about any other direction but ${\bf n}$. This is simply because the very meaning
of a particular bivector ${{\boldsymbol\mu}\cdot{\bf n}}$ is defined by the direction ${\bf n}$.

It should be clear from the above picture that the outcome of a measurement of the
observable ${A_{\bf n}(\boldsymbol\mu)}$ cannot yield anything but the binary values
${\pm\,1}$. It will yield ${+\,1}$ if the bivector ${{\boldsymbol\mu}\cdot{\bf n}}$ is
spinning counterclockwise about the direction ${\bf n}$, and ${-\,1}$ if the
bivector is spinning clockwise about ${\bf n}$. Moreover, it will yield null result (i.e.,
zero) if the spin analyzer is not aligned with the direction ${\bf n}$ of the bivector.
This is because a projection onto (or inner product with) the direction ${\bf n}$ of any
arbitrary bivector different from ${{\boldsymbol\mu}\cdot{\bf n}}$ would not yield another
bivector but an ordinary vector, in the direction {\it orthogonal} to the original direction
${\bf n}$ (cf. pp ${30-33}$ of Ref.\ocite{Doran}). It is also worth stressing here that
the above picture is not meant to suggest that spin is somehow ``created'' by the spin
analyzer, but simply that only the component ${{\boldsymbol\mu}\cdot{\bf n}}$ of the 
trivector ${\boldsymbol\mu}$---obtained by a projection along the direction ${\bf n}$ of
the analyzer---will pass through that analyzer. And when this particular component of
${\boldsymbol\mu}$ does pass through, it can only trigger either the ``spin up'' detector
or the ``spin down'' detector, for the resulting bivector ${{\boldsymbol\mu}\cdot{\bf n}}$
could only be spinning ``up'' or ``down'' along the direction ${\bf n}$. Thus, when probed
individually, no scrutiny by means of a spin analyzer can extract anything but a {\it bit}
of information from the observables ${A_{\bf n}(\boldsymbol\mu)}$, since, apart from the
sense of their rotation, no other properties of these observables are unspecified.
It should now be fairly clear how the binary value assignment of these observables, along
with the value {\it zero} for a possible null result, comes about. It is, in fact, no
different from that illustrated
by Clauser and Shimony in Figure 1 of their celebrated report \ocite{Clauser-Shimony}.
Thus, our observables ${A_{\bf a}(\boldsymbol\mu)}$ and ${B_{\bf b}(\boldsymbol\mu)}$
represent quite faithfully what is actually observed in a Stern-Gerlach type experiment
(see also Ref.\ocite{reply}). 

To be sure, a diehard nonlocalist may continue to see here an opportunity to dissent. For,
although our dichotomic observables ${A_{\bf a}(\boldsymbol\mu)}$ and
${B_{\bf b}(\boldsymbol\mu)}$ can only yield the binary values ${\pm\,1}$ as demanded by
quantum mechanical predictions, there still remains a psychological problem of seeing
bivectors where one is used to seeing commuting real numbers. It would be a naive mistake,
however, to read too much into this bivectorial representation of the observable physical
quantities. To begin with, as we have already noted, the perspective on the relation between
scalars and bivectors is radically different in Clifford algebra from the conventional
perspective fostered by the vector algebra. In particular, scalars, vectors, bivectors, and
trivectors are all intimately linked in Clifford algebra by the invertible geometric product
of almost magical significance. What is more, the bivectors in the Clifford algebra ${Cl_{3,0}}$
are in fact isomorphic to the familiar quaternionic numbers. More precisely, left-handed set of
orthonormal bivectors is isomorphic to the right-handed set of pure quaternionic numbers
\ocite{Doran}. Consequently, the observable physical quantities within our model can
be equivalently thought of as either ``real bivectors'' or ``complex quaternions''. In other
words, there is nothing intrinsically either bivectorial or complex about the observables
${A_{\bf a}(\boldsymbol\mu)}$ and ${B_{\bf b}(\boldsymbol\mu)}$. These are simply their
representation-dependent features, and hence cannot be expected to have deeper physical
significance. The feature that remains truly independent of a particular representation is
the {\it non-commutativity} of their products, and hence it is this non-commutativity that
has a genuine physical significance. It would play a decisive role in what follows.

Is it, however, reasonable to use non-commuting numbers such as quaternions in a local
realistic theory? Of course it is. Aerospace engineers routinely use non-commuting
quaternions in applications to rotations in the ordinary Euclidean space, precisely because
they do not commute. Moreover, this lack of commutativity of quaternions merely reflects the fact
that the basis vectors of Euclidean space can be chosen to be orthogonal---a fact of geometry,
not of dynamics \ocite{Doran}. Thus, {\it a priori}, the local realists are by no means obliged
to remain unimaginative and consider only the commuting real numbers for their theories.
Furthermore, let us not forget---as so often unguarded preoccupation
with operationalism makes us do---that physically
the observables ${A_{\bf a}(\boldsymbol\mu)}$ and ${B_{\bf b}(\boldsymbol\mu)}$ are supposed to
have more significance than just representing the results of spin measurements. They must, in
fact, also represent the {\it dynamical variables} of a yet to be discovered physical theory,
which, when measured, should reproduce the binary outcomes ${\pm\,1}$. That is to say, the
observables ${A_{\bf a}(\boldsymbol\mu)}$ and ${B_{\bf b}(\boldsymbol\mu)}$ are supposed
to be the dispersion-free counterparts of the quantum mechanical spin operators
${{\boldsymbol\sigma}\cdot{\bf a}}$ and ${{\boldsymbol\sigma}\cdot{\bf b}}$, in addition to
having operational significance of representing the
measurement outcomes  ${\pm\,1}$ \ocite{Bell-1966}.
Indeed, as Bell himself emphasized: ``In a complete physical theory of the type envisaged by
Einstein, the hidden variables would have dynamical significance and laws of
motion...''${\,}$\ocite{Bell-1964}. If,
then, one of these hidden variables is allowed to be a Clifford algebra valued variable
${\boldsymbol\mu}$ (and this much the nonlocalists must allow if they are to remain in the
game), then, inevitably, the resulting observables ${A_{\bf a}({\boldsymbol\mu})}$
and ${B_{\bf b}({\boldsymbol\mu})}$, in addition to having operational attributes, would also
inherit the Clifford-algebraic attributes from ${\boldsymbol\mu}$, and hence be non-commuting.  
{\it A priori}, then, there seems to be no reason for excluding non-commuting observables from
all conceivable local realistic theories, so long as they do not contradict any other
premises of Bell's theorem, or the predictions of quantum mechanics.

Could it be, however, that one of the premises of Bell's theorem, such as locality, actually
precludes non-commuting observables from consideration, if only {\it a posteriori}${\,}$? As we
shall soon see, the answer to this question is: {\it No${\,}$!} At least not {\it
counterfactually} ${\,}$(in a sense that will become clear soon). In the following section---by
proving the conditions of remote parameter independence and remote outcome independence
\ocite{Shimony-2004}---we shall explicitly show that the model constructed in
Ref.\ocite{Christian} is strictly locally causal, even at the level of the individual,
uncontrollable, microstates.

\section{Local causality of the model at the level of microstates}

In the original formulation of his theorem \ocite{Bell-1964}, Bell considered a joint spin
observable, such as the product ${(A_{\bf a}\,B_{\bf b})}$ in our notation, which is a single
observable of an EPR-Bohm type pair of particles, but requiring two distinct operations for
its measurement. Since in a deterministic hidden variable theory this observable should
have a definite value, say ${(A_{\bf a}\,B_{\bf b})(\lambda)}$, for each microstate
${\lambda}$, Bell required the joint observable ${(A_{\bf a}\,B_{\bf b})}$ to satisfy the
locality condition \ocite{Clauser-Shimony}
\begin{equation}
(A_{\bf a}\,B_{\bf b})(\lambda)\,=\,
A_{\bf a}(\lambda)\,B_{\bf b}(\lambda)\,,\label{factori-2}
\end{equation}
which states that the value of the product observable ${(A_{\bf a}\,B_{\bf b})}$ is necessarily
equal to the product of the values of the two individual observables ${A_{\bf a}}$ and
${B_{\bf b}}$, if the
hidden variable theory in question is to be locally causal in addition to being realistic and
deterministic. Moreover, it is evident from the notation used in the above equation that, once
the microstate ${\lambda}$ is specified and the particles have separated, measurement outcomes
of the local observable ${A_{\bf a}}$ do not depend on the remote parameter
${\bf b}$, but only upon the microstates ${\lambda}$ and the local parameter ${\bf a}$, and
likewise for the observable ${B_{\bf b}}$. In other words, the above factorizability condition
requires that the actual values of the three observables ${A_{\bf a}}$, ${B_{\bf b}}$, and
${(A_{\bf a}\,B_{\bf b})}$---for all ${\bf a}$, ${\bf b}$, and ${\lambda}$---must satisfy
the following relationship among themselves:
\begin{equation}
v(A,\,B)\,=\,A\,B\,,\label{loc-1}
\end{equation}
where, respectively, ${A=-1}$, ${0}$, or ${+1}$, and ${B=-1}$, ${0}$, or ${+1}$ are
the values that could be found if the observables ${A_{\bf a}}$ and
${B_{\bf b}}$ are measured in some experiment, and ${v(A,\,B)}$ is
the corresponding value of the joint observable ${(A_{\bf a}\,B_{\bf b})}$.

For any deterministic local hidden variable theory, the factorizability condition
(\ref{factori-2}) stated above is both necessary and sufficient to guarantee the local
causality of the theory. Therefore our adaptation in Ref.\ocite{Christian} of Bell's
original local realistic framework based on this condition is perfectly adequate.
Unfortunately, the apparent non-commutativity of our observables has raised suspicion
that perhaps this condition is not satisfied within our model after all, and perhaps
there is some subtle form of nonlocality lurking beneath the surface, especially
because the model reproduces the quantum mechanical correlations so exactly. Such a
suspicion, however, as we shall soon see, is without merit.

To be sure, the non-commutativity of observables plays a central role in our model, and
it is indeed hiding in the equation (\ref{bi-identity}) above (or in the equation (17)
of Ref.\ocite{Christian}). It can be made explicit by writing
\begin{equation}
[\,{\boldsymbol\mu}\cdot{\bf a},\,{\boldsymbol\mu}\cdot{\bf b}\,]\,=
\,-\,2\,{\boldsymbol\mu}\cdot({\bf a}\times{\bf b})\,=\,
\,-\,2\,({\boldsymbol\mu}\cdot{\bf z})\,\sin\theta_{{\bf a}{\bf b}}\,,\label{nonecome}
\end{equation}
where ${\theta_{{\bf a}{\bf b}}}$ is the angle from ${\bf a}$ to ${\bf b}$ about the
unit direction ${{\bf z}\equiv({\bf a}\times{\bf b})/\sin\theta_{{\bf a}{\bf b}}\,}$.
Admittedly, at first sight such a blatant non-commutativity does seem to be in
conflict with the locality condition (\ref{loc-1}), since this condition implies
\begin{equation}
v(A,\,B)\,-\,v(B,\,A)\,=\,A\,B\,-\,B\,A\,=\,[\,A,\,B\,]\,=\,0.\label{loc-commu}
\end{equation}
This apparent conflict evaporates, however, as soon as it is appreciated that the numbers ${A}$
and ${B}$ here are the {\it actual} values of the observables ${{\boldsymbol\mu}\cdot{\bf a}}$ 
and ${{\boldsymbol\mu}\cdot{\bf b}}$, whereas the observable ${{\boldsymbol\mu}\cdot{\bf z}}$
appearing on the RHS of the equation (\ref{nonecome}) can contribute to physics
only {\it counterfactually}.
This is because the direction ${\bf z}$ defining the observable ${{\boldsymbol\mu}\cdot{\bf z}}$
is orthogonal to both ${\bf a}$ and ${\bf b}$, and hence it is necessarily {\it exclusive} to at
least one of them. Consequently, in any EPR-Bohm type experiment, simultaneous measurements of
either the pair ${\{{\boldsymbol\mu}\cdot{\bf a}}$,\;${{\boldsymbol\mu}\cdot{\bf z}\}}$ or the
pair ${\{{\boldsymbol\mu}\cdot{\bf b}}$,\;${{\boldsymbol\mu}\cdot{\bf z}\}}$ of observables
would be {\it impossible}. That is to say, a measurement of any one member of one of the pairs
at a chosen station would preclude the measurement of the second member of the same pair at
the same station. This, in turn, means that simultaneous measurements of the observables
${{\boldsymbol\mu}\cdot{\bf a}}$, ${{\boldsymbol\mu}\cdot{\bf b}}$, and
${{\boldsymbol\mu}\cdot{\bf z}}$, even using both stations, would require experimental
arrangements along at least two mutually exclusive directions, rendering the joint measurement
of all three of them impossible. Simply put, as far as we know, a spin analyzer at a given station
cannot be aligned to two mutually exclusive directions at the same time. And even if the
inhabitants of some advanced civilization may manage to align their analyzers along two mutually
exclusive Euclidean directions, the detector along one of these two directions would not respond,
because the second particle would be near the remote end of the EPR experiment. In other words,
for a given pair of particles, whenever the two observables, say ${{\boldsymbol\mu}\cdot{\bf a}}$
and ${{\boldsymbol\mu}\cdot{\bf b}}$, are found to have non-zero values, such as the values ${A}$
and ${B}$ above, the only value the third observable ${{\boldsymbol\mu}\cdot{\bf z}}$ could
possibly be found to have is {\it zero}, corresponding to non-detection, regardless of the
station that could have been used to measure it. Thus, for any physically realizable experiment 
the RHS of the equation (\ref{nonecome}) would necessarily vanish. Consequently, in any actual
experiment, the commutativity condition (\ref{loc-commu}) is necessarily satisfied within our
model, despite the manifest non-commutativity of our observables. Counterfactually, on the other
hand, the RHS of the equation (\ref{nonecome}) plays a crucial role in generating the EPR-Bohm
correlations within our model, as we shall see in section IV below. 

Together with the analysis in Ref.\ocite{Christian}, the above argument is adequate
enough to reaffirm the local causality of our model. A deterministic, local realistic model
such as ours, however, should be able to do better than relying on an operationally flavored
argument for its consistency. Let us, therefore, investigate exactly how the local causality
is maintained within our model, independently of what any experimental constraint forbids us
to do. For this purpose, it is convenient to analyze the equation
(\ref{nonecome}) above in terms of the conditions of remote parameter independence and
remote outcome independence \ocite{Shimony-2004}. It is well known that Bell's locality
condition (\ref{factori-2}) is equivalent to the conjunction of these two conditions. More
significantly, these conditions allow us to demonstrate that the non-commutativity of the
observables within our local model is quite harmless, and reflects merely the intrinsic
geometrical features of rotations in the physical space.
To this end, let us first focus on the condition of remote parameter independence.

\subsection{Remote Parameter Independence within the Local Model}

The deterministic version of this condition states that, for a given microstate ${\boldsymbol\mu}$,
the outcome of an experiment at a local station {\bf 1} must not dependent on the chosen
experimental settings at the remote station {\bf 2}, and vice versa. Now, our model is a
{\it deterministic} hidden variable model. Nevertheless, it is convenient to state the
{\it probabilistic} version of this condition in a symbolic form, because it makes the
concepts involved more transparent:
\begin{align}
P_{\boldsymbol\mu}(A\,|\,{\bf a},\,{\bf b'})\,&=\;
P_{\boldsymbol\mu}(A\,|\,{\bf a},\,{\bf b}) \label{symbolpara} \\
{\rm and}\;\;\;P_{\boldsymbol\mu}(B\,|\,{\bf a'},\,{\bf b})\,&=\;
P_{\boldsymbol\mu}(B\,|\,{\bf a},\,{\bf b}),\;\;\;\;\;\;\;\;\;\;
\end{align}
where ${\bf a}$ and ${\bf a'}$ and ${\bf b}$ and ${\bf b'}$ are mutually exclusive directions
at stations {\bf 1} and {\bf 2}, respectively. The question we now wish to address is whether
the deterministic version of these conditions hold within our local model, despite the 
non-commutativity of observables expressed in equation
(\ref{nonecome}). To answer this question, let us rewrite equation (\ref{nonecome}) as
\begin{equation}
[A_{\bf a}(\boldsymbol\mu),\,B_{\bf b}(\boldsymbol\mu)]\,=
\,-\,2\;{\boldsymbol\mu}\cdot({\bf a}\times{\bf b})\,,\label{obsnonecome}
\end{equation}
and use the fact that the Clifford product is invertible in general. This allows us to express
the equation as
\begin{equation}
A_{\bf a}(\boldsymbol\mu)\,=\,B_{\bf b}(\boldsymbol\mu)\,A_{\bf a}(\boldsymbol\mu)\,
B^{-1}_{\bf b}(\boldsymbol\mu)\,-\,2\;\{{\boldsymbol\mu}\cdot({\bf a}\times{\bf b})\}\,
B^{-1}_{\bf b}(\boldsymbol\mu)\,.\label{noseobsno}
\end{equation}

At first sight, it appears from this relation that---within our model---a remote observable
such as ${B_{\bf b}({\boldsymbol\mu})}$ can indeed influence a local observable
${A_{\bf a}({\boldsymbol\mu})}$, and vice versa. In fact, despite the manifest functional
independence from ${\bf b}$ of the observable ${A_{\bf a}({\boldsymbol\mu})}$, it appears
that even the choice of a remote parameter ${\bf b}$ can affect
${A_{\bf a}({\boldsymbol\mu})}$ in a direct manner, thereby violating the basic principles
of relativity. This, however, is not the case. Despite appearances, the observable
${A_{\bf a}({\boldsymbol\mu})}$ cannot be influenced by either the remote parameter ${\bf b}$
or the remote outcome ${B}$. In fact, the above relation is simply a geometrical statement
expressing exactly the opposite. In particular, it asserts that any changes in the parameter
${\bf b}$ on the RHS of equation (\ref{noseobsno})---such as from
${\bf b}$ to ${\bf b'}$---cannot affect the observable ${A_{\bf a}(\boldsymbol\mu)}$ on its
LHS. The situation is analogous to that in gauge invariance: A simultaneous change of ``gauge''
in the two ``gauge-dependent'' terms on the RHS of equation (\ref{noseobsno}) does not affect
their ``gauge invariant'' sum on the LHS. To appreciate this fact, let us note that the
condition of remote parameter independence can be said to hold in our model if the LHS of
equation (\ref{noseobsno}) can be shown to be unaffected by the changes in the parameter
${\bf b}$ on its RHS, in line with the condition (\ref{symbolpara}) above. That is to say,
remote parameter independence is said to hold in our model if the following equality holds:
\begin{equation}
B_{\bf b'}(\boldsymbol\mu)\,A_{\bf a}(\boldsymbol\mu)\,
B^{-1}_{\bf b'}(\boldsymbol\mu)\,-\,2\,\{{\boldsymbol\mu}\cdot({\bf a}\times{\bf b'})\}\,
B^{-1}_{\bf b'}(\boldsymbol\mu)\,=\,
B_{\bf b}(\boldsymbol\mu)\,A_{\bf a}(\boldsymbol\mu)\,
B^{-1}_{\bf b}(\boldsymbol\mu)\,-\,2\,\{{\boldsymbol\mu}\cdot({\bf a}\times{\bf b})\}\,
B^{-1}_{\bf b}(\boldsymbol\mu).\label{whjateobsno}
\end{equation}
This apparently complicated equality can be greatly simplified by using the relation
${B^{-1}_{\bf b}(\boldsymbol\mu)\,=\,-\,B_{\bf b}(\boldsymbol\mu)}$ (which is true for any
bivector), and then substituting
the explicit expressions for the observables appearing in it. This gives
\begin{equation}
-\,{\boldsymbol\mu}\,{\bf b'}\,{\boldsymbol\mu}\,{\bf a}\,{\boldsymbol\mu}\,{\bf b'}\,+\,2\,
{\boldsymbol\mu}({\bf a}\times{\bf b'})\,{\boldsymbol\mu}{\bf b'}\,=\,
-\,{\boldsymbol\mu}\,{\bf b}\,{\boldsymbol\mu}\,{\bf a}\,{\boldsymbol\mu}\,{\bf b}\,+\,2\,
{\boldsymbol\mu}({\bf a}\times{\bf b})\,{\boldsymbol\mu}{\bf b}\,.
\end{equation}
By elementary manipulations and a use of the triple cross product identity, this equality
can be further reduced to
\begin{equation}
{\boldsymbol\mu}\{{\bf b'}\,{\bf a}\,{\bf b'}\,-\,2\,{\bf b'}({\bf a}\cdot{\bf b'})\,+\,2\,
{\bf a}\}\,=\,
{\boldsymbol\mu}\{{\bf b}\,{\bf a}\,{\bf b}\,-\,2\,{\bf b}({\bf a}\cdot{\bf b})\,+\,2\,
{\bf a}\}\,,
\end{equation}
which---{\it nota bene}---involves Clifford products of ordinary vectors. 
Now this equality would hold if the relation
\begin{equation}
{\bf b'}\,{\bf a}\,{\bf b'}\,-\,2\,{\bf b'}({\bf a}\cdot{\bf b'})\,=\;
{\bf b}\,{\bf a}\,{\bf b}\,-\,2\,{\bf b}({\bf a}\cdot{\bf b})
\end{equation}
is satisfied. But this relation is indeed satisfied, because both sides of it are simply
elementary geometric expressions of one and the same vector in the Euclidean space. To
appreciate this, first note
that ${{\bf b}\,{\bf a}\,{\bf b}}$ is simply a reflection of ${\bf a}$ across ${\bf b}$, in
the plane defined by ${\bf a}$ and ${\bf b}$, and ${{\bf b}\,({\bf a}\cdot{\bf b})}$ is the
projection of ${\bf a}$ along ${\bf b}$ (cf. Fig.${\,}$9 of Ref.\ocite{Oersted}). It is then
easy to realize that the above equality does hold, with both sides being simply equal to
${-\,{\bf a}}$. Consequently, the equality (\ref{whjateobsno}) also holds, and hence the
remote parameter independence within our model is indeed satisfied.

\subsection{Remote Outcome Independence within the Local Model}

But perhaps the remote parameter independence of the model was never in doubt. It is fairly
obvious from the bivectorial definition of the observables ${A_{\bf a}({\boldsymbol\mu})}$
and ${B_{\bf b}({\boldsymbol\mu})}$ that they have nothing to do with the remote contexts
${\bf b}$ and ${\bf a}$, respectively. Perhaps, then, the model is nonlocal in a more subtle
manner, and violates somehow the condition of remote {\it outcome} independence, in a manner
reminiscent of
quantum mechanics. After all, it does reproduce the relevant predictions of quantum mechanics
{\it exactly}. We shall soon see, however, that, on the contrary, our model strictly respects
remote outcome independence, and does so unequivocally. To appreciate this fact, let us
first recall the precise meaning of the condition of remote outcome independence. The
deterministic version of this condition states that, for a given microstate ${\boldsymbol\mu}$,
the outcome of an experiment at a local station {\bf 1} must not dependent on the outcome at 
the remote station {\bf 2}, and vice versa. Again, for the reasons of conceptual transparency,
despite the fact that ours is a {\it deterministic} hidden variable model, we restate
this condition symbolically in its {\it probabilistic} form:
\begin{align}
P_{\boldsymbol\mu}(A=+1\,|\,{\bf a},\,{\bf b},\,B=+1)\,&=\;
P_{\boldsymbol\mu}(A=+1\,|\,{\bf a},\,{\bf b},\,B=-1)\,, \label{symbolprob-1} \\
P_{\boldsymbol\mu}(A=-1\,|\,{\bf a},\,{\bf b},\,B=+1)\,&=\;
P_{\boldsymbol\mu}(A=-1\,|\,{\bf a},\,{\bf b},\,B=-1)\,, \label{symbolprob-2}
\end{align}
and similar equalities with ${A}$ and ${B}$ interchanged. It is worth emphasizing here that
this condition does not preclude correlations between the outcomes ${A}$ and ${B}$ at the
two ends of an EPR experiment. Rather, it asserts that, given a complete state
${\boldsymbol\mu}$, the outcome at one end of the experiment provides no {\it additional}
information concerning the outcome at the other end, and vice versa \ocite{Shimony-2004}.
Again, ${\,}$the question we wish to address here is whether the deterministic versions of the
above equalities hold within our local model. To answer this question, let us rewrite
equation (\ref{nonecome}) again as
\begin{equation}
A_{\bf a}(\boldsymbol\mu)\,=\,-\,B_{\bf b}(\boldsymbol\mu)\,A_{\bf a}(\boldsymbol\mu)\,
B_{\bf b}(\boldsymbol\mu)\,+\,2\;C_{\bf z}(\boldsymbol\mu)\,
B_{\bf b}(\boldsymbol\mu)\,\sin\theta_{{\bf a}{\bf b}}\,,\label{obsnonec}
\end{equation}
where ${C_{\bf z}(\boldsymbol\mu):={\boldsymbol\mu}\cdot{\bf z}\,}$, and we have again used
the fact that ${B^{-1}_{\bf b}(\boldsymbol\mu)\,=\,-\,B_{\bf b}(\boldsymbol\mu)}$.
As we have already noted, at first sight this relation seems to imply that the measurement
outcomes of the remote observable ${B_{\bf b}({\boldsymbol\mu})}$ can influence those of
the local observable ${A_{\bf a}({\boldsymbol\mu})}$, and vice versa. Again, as a first
step towards proving the contrary, we observe that the condition of remote outcome
independence can be said to hold in our model if the LHS of equation (\ref{obsnonec})
can be shown to be unaffected by the changes in the measurement outcomes of the observable
${B_{\bf b}(\boldsymbol\mu)}$ on its RHS, in line with, say, the condition (\ref{symbolprob-1})
above. That is to say, remote outcome independence is said to hold in our model if the equality
\begin{equation}
-\,B^{(+)}_{\bf b}(\boldsymbol\mu)\,A^{(+)}_{\bf a}(\boldsymbol\mu)\,
B^{(+)}_{\bf b}(\boldsymbol\mu)\,+\,2\;C^{(+)}_{\bf z}(\boldsymbol\mu)\,
B^{(+)}_{\bf b}(\boldsymbol\mu)\,\sin\theta_{{\bf a}{\bf b}}\,=
-\,B^{(-)}_{\bf b}(\boldsymbol\mu)\,A^{(+)}_{\bf a}(\boldsymbol\mu)\,
B^{(-)}_{\bf b}(\boldsymbol\mu)\,+\,2\;C^{(-)}_{\bf z}(\boldsymbol\mu)\,
B^{(-)}_{\bf b}(\boldsymbol\mu)\,\sin\theta_{{\bf a}{\bf b}}\label{outindi}
\end{equation}
holds, together with the three other analogous equalities. Here the signs in parentheses
over the observables indicate whether the corresponding bivector is rotating counterclockwise
${(+)}$ or clockwise ${(-)}$. Clearly, the signs over the observables
${A_{\bf a}({\boldsymbol\mu})}$ and ${B_{\bf b}({\boldsymbol\mu})}$ are determined by the
choice of the equality, but it may not be obvious how the signs over the third
observable ${C_{\bf z}({\boldsymbol\mu})}$ have come about. They are, in fact, a result of
the geometrical fact that the bivector corresponding to ${C_{\bf z}({\boldsymbol\mu})}$
would be rotating counterclockwise ${(+)}$ when the other two bivectors are rotating in the
same sense (either both counterclockwise or both clockwise), and it would be rotating
clockwise ${(-)}$ when the other two are rotating in the opposite senses to each other. This
is because ${{\boldsymbol\mu}\cdot{\bf z}}$ is not an independent bivector, but defined
by the direction resulting from the cross product of the directions defining the other two
bivectors. Thus, for example, when ${A_{\bf a}({\boldsymbol\mu})}$ happens to be
equal to ${(+I)(+{\bf a})}$ and ${B_{\bf b}({\boldsymbol\mu})}$ happens to be
equal to ${(-I)(+{\bf b})=(+I)(-{\bf b})}$, then ${C_{\bf z}({\boldsymbol\mu})}$ would clearly
be equal to ${(+I)(-{\bf z})=(-I)(+{\bf z})}$, and hence it would be rotating in the clockwise
sense about the direction ${+{\bf z}}$. In short, when ${{\boldsymbol\mu}\cdot{\bf a}}$ and
${{\boldsymbol\mu}\cdot{\bf b}}$ are both spinning ``up'', then
${{\boldsymbol\mu}\cdot{\bf z}}$ would be spinning ``up'' as well, but when, say,
${{\boldsymbol\mu}\cdot{\bf a}}$ is spinning ``up'' and
${{\boldsymbol\mu}\cdot{\bf b}}$ is spinning ``down'', then ${{\boldsymbol\mu}\cdot{\bf z}}$
would be spinning down, along ${+{\bf z}\,}$, and so on.

Once these geometrical facts are appreciated, it is simply a matter of counting the signs to
realize that the above equality does hold within our model, and so do the other three
equalities. This proves that the observable ${A_{\bf a}({\boldsymbol\mu})}$ and its
measurement outcomes on the LHS of equation (\ref{obsnonec}) are unaffected, not only
by the measurement outcomes of the remote observable ${B_{\bf b}({\boldsymbol\mu})}$, but
also by whether or not the latter is actually measured \ocite{Einstein-1948}. This, in turn,
means that the condition of remote outcome independence
also holds within our model. What is more, by simultaneously changing the parameter
${\bf b}$ and the outcome ${B}$ in the equation 
(\ref{obsnonec}) above, it is easy to prove that the two independence conditions hold within our
model not only separately, but also in conjunction. This, then, completes the purely
geometric proof of the local causality of our model, at the level of individual
microstates. As an immediate implication, the model allows a common cause
explanation of the EPR-Bohm correlations, as we now proceed to demonstrate.

\section{Statistical interpretation of the entangled singlet state}

As is well known, Einstein wished to interpret the entangled quantum state as describing
an ensemble of micro systems rather than an individual quantum system \ocite{Einstein}.
Such a statistical
interpretation of the entangled quantum state would be justified if a quantitatively precise
account of the predictions based on such a state can be provided, in terms of an ensemble
of sub-quantum microstates. Here we shall provide such an account for the entangled singlet
state. To be sure, the derivations of the equation (19)
of Ref.\ocite{Christian} (or of the equation (5) of Ref.\ocite{reply}) may well thought to
be sufficient for such an account, but it is instructive to see how well the account holds
up when extended to as intricate a scenario as an actual CHSH type experiment \ocite{Clauser}.
We shall see that the non-commutativity of observables displayed in the equation
(\ref{nonecome}) above plays a crucial role in providing the quantitative aspects of this account.

To this end, let us reconsider the familiar CHSH string of
expectation values \ocite{Clauser-Shimony}:
\begin{equation}
{\cal E}({\bf a},\,{\bf b})\,+\,{\cal E}({\bf a},\,{\bf b'})\,+\,
{\cal E}({\bf a'},\,{\bf b})\,-\,{\cal E}({\bf a'},\,{\bf b'})\,.
\label{CHSH-op}
\end{equation}
For the microstates ${\boldsymbol\mu}$ of our local model this string can be
rewritten using the notations of Ref.\ocite{Christian} as
\begin{equation}
{\cal E}({\bf a},\,{\bf b})\,+\,{\cal E}({\bf a},\,{\bf b'})\,+\,
{\cal E}({\bf a'},\,{\bf b})\,-\,{\cal E}({\bf a'},\,{\bf b'})\;=\,
\int_{{\cal V}_3}{\cal F}_{c.v.}(\boldsymbol\mu)\;\,
d{\boldsymbol\rho}(\boldsymbol\mu)\,, \label{probint}
\end{equation}
with the corresponding local realistic function ${{\cal F}_{c.v.}(\boldsymbol\mu)}$ defined as
\begin{equation}
{\cal F}_{c.v.}(\boldsymbol\mu):=A_{\bf a}(\boldsymbol\mu)
\left\{\,B_{\bf b}(\boldsymbol\mu)+B_{\bf b'}(\boldsymbol\mu)\,\right\}
+A_{\bf a'}(\boldsymbol\mu)\left\{\,B_{\bf b}(\boldsymbol\mu)
-B_{\bf b'}(\boldsymbol\mu)\,\right\}\!.
\end{equation}
If we now use the fact that the observables ${A^2_{\bf a}(\boldsymbol\mu)}$,
${A^2_{\bf a'}(\boldsymbol\mu)}$, ${B^2_{\bf b}(\boldsymbol\mu)}$, and
${B^2_{\bf b'}(\boldsymbol\mu)}$ are all equal to ${-1}$ (because they are unit bivectors),
then the square of the function ${{\cal F}_{c.v.}(\boldsymbol\mu)}$ simplifies to 
(cf. Ref.\ocite{bounds})
\begin{equation}
{\cal F}_{c.v.}^{\,2}(\boldsymbol\mu)\,=\,4\,+\,\left[\,A_{\bf a}(\boldsymbol\mu),\,
A_{\bf a'}(\boldsymbol\mu)\,\right]\left[\,B_{\bf b'}(\boldsymbol\mu),\,
B_{\bf b}(\boldsymbol\mu)\,\right]\!,\label{20/20}
\end{equation}
provided we assume that both ${A}$'s commute with both ${B}$'s, and vice versa:
\begin{equation}
\left[\,A_{\bf n}(\boldsymbol\mu),\,B_{\bf n'}(\boldsymbol\mu)\,\right]\,=\,0\,,
\;\;\;\forall\;\,{\bf n}\;\,{\rm and}\;\,{\bf n'}.
\end{equation}
Using equation (18) of Ref.\ocite{Christian} it is easy to see that the
vanishing of the above commutator does hold for our observables
${A_{\bf a}(\boldsymbol\mu)}$ and ${B_{\bf b}(\boldsymbol\mu)}$, at least
upon averaging over ${\boldsymbol\mu}$, which is sufficient
for the purposes of this section. Next, using the identity (\ref{bi-identity}),
the Clifford product of the local commutators appearing in equation (\ref{20/20})
can be worked out as\footnotemark\footnotetext{Equations (\ref{mikee}) and
(\ref{see-0}) were first worked out by Michael Seevinck in a private
correspondence regarding Ref.\ocite{Christian}.}
\begin{equation}
\left[\,A_{\bf a}(\boldsymbol\mu),\,
A_{\bf a'}(\boldsymbol\mu)\,\right]\left[\,B_{\bf b'}(\boldsymbol\mu),\,
B_{\bf b}(\boldsymbol\mu)\,\right]\,
=\,4\;[\,{\boldsymbol\mu}\cdot({\bf a}\times{\bf a'})]\,
[\,{\boldsymbol\mu}\cdot({\bf b'}\times{\bf b})].\label{mikee}
\end{equation}
Averaging this product over the uncontrollable microstates ${\boldsymbol\mu}$
(just as in equation (19) of Ref.\ocite{Christian}) then gives
\begin{equation}
\int_{{\cal V}_3}\left[\,A_{\bf a}(\boldsymbol\mu),\,
A_{\bf a'}(\boldsymbol\mu)\,\right]\left[\,B_{\bf b'}(\boldsymbol\mu),\,
B_{\bf b}(\boldsymbol\mu)\,\right]\,d{\boldsymbol\rho}(\boldsymbol\mu)\,
=\,-\,4\,({\bf a}\times{\bf a'})\cdot({\bf b'}\times{\bf b})\,.\label{see-0}
\end{equation}
This relation is of course equivalent to averaging over both sides of the
equation (\ref{20/20}):
\begin{equation}
\int_{{\cal V}_3}{\cal F}^{\,2}_{c.v.}(\boldsymbol\mu)\;\,
d{\boldsymbol\rho}(\boldsymbol\mu)\,=\,4\,+\!
\int_{{\cal V}_3}\left[\,A_{\bf a}(\boldsymbol\mu),\,
A_{\bf a'}(\boldsymbol\mu)\,\right]\left[\,B_{\bf b'}(\boldsymbol\mu),\,
B_{\bf b}(\boldsymbol\mu)\,\right]\,d{\boldsymbol\rho}(\boldsymbol\mu)\,
=\,4\,-\,4\,({\bf a}\times{\bf a'})\cdot({\bf b'}\times{\bf b}).
\end{equation}
From this average, using the variance inequality
\begin{equation}
\left|\int_{{\cal V}_3}{\cal F}_{c.v.}(\boldsymbol\mu)\;\,
d{\boldsymbol\rho}(\boldsymbol\mu)\right|^2\,\leq\,
\int_{{\cal V}_3}{\cal F}^{\,2}_{c.v.}(\boldsymbol\mu)\;\,
d{\boldsymbol\rho}(\boldsymbol\mu)\,, \label{rms}
\end{equation}
we finally arrive at the violations of the CHSH inequality within our local model:
\begin{equation}
|\;{\cal E}({\bf a},\,{\bf b})\,+\,{\cal E}({\bf a},\,{\bf b'})\,+\,
{\cal E}({\bf a'},\,{\bf b})\,-\,{\cal E}({\bf a'},\,{\bf b'})\,|\,\leq\,
\sqrt{4\,+\,4\;|\,({\bf a}\times{\bf a'})\cdot({\bf b'}\times{\bf b})\,|\,}\,
\leq\,2\sqrt{2}\,.\label{see-2}
\end{equation}
This result is in quantitatively precise agreement with the corresponding prediction
of quantum mechanics \ocite{bounds}\ocite{Mike}. That there is such a precise agreement
between the prediction of our local model and that of quantum mechanics was
already demonstrated in Ref.\ocite{reply}. To appreciate it further, let us calculate the
expectation value in the singlet state of the quantum mechanical
analogue of the equation (\ref{20/20}), by means of the so-called Bell operator:
\begin{equation}
{\cal B}_{op}\,=\,
{\boldsymbol\sigma}_1\cdot{\bf a}\otimes{\boldsymbol\sigma}_2\cdot{\bf b}\,+\,
{\boldsymbol\sigma}_1\cdot{\bf a}\otimes{\boldsymbol\sigma}_2\cdot{\bf b'}\,+\,
{\boldsymbol\sigma}_1\cdot{\bf a'}\otimes{\boldsymbol\sigma}_2\cdot{\bf b}\,-\,
{\boldsymbol\sigma}_1\cdot{\bf a'}\otimes{\boldsymbol\sigma}_2\cdot{\bf b'}.
\end{equation}
The square of this operator is of course well known \ocite{bell-op},
and works out to be
\begin{equation}
{\cal B}_{op}^{\,2}\,=\,4\,\dbl\,+\,4\,\{{\boldsymbol\sigma}_1\cdot({\bf a}\!\times\!{\bf a'})\}
\otimes\{{\boldsymbol\sigma}_2\cdot({\bf b}\!\times\!{\bf b'})\}
\end{equation}
(where ${\dbl}$ is the identity operator), with its expectation value in the entangled
singlet state being
\begin{equation}
\langle\,{\cal B}_{op}^{\,2}\,\rangle\,=\,4\,+\,4\,\left\langle\,
\{{\boldsymbol\sigma}_1\cdot({\bf a}\!\times\!{\bf a'})\}
\otimes\{{\boldsymbol\sigma}_2\cdot({\bf b}\!\times\!{\bf b'})\}\,\right\rangle\,=\,
4\,-\,4\,({\bf a}\times{\bf a'})\cdot({\bf b}\times{\bf b'}).
\end{equation}
Once again using the variance inequality, we arrive at the quantum mechanical violations of
the CHSH inequality \ocite{bounds}:
\begin{equation}
|\,\langle\,{\cal B}_{op}\,\rangle\,|\,\leq\,
\sqrt{4\,+\,4\;|\,({\bf a}\times{\bf a'})\cdot({\bf b}\times{\bf b'})\,|\,}\,\leq\,2\sqrt{2}\,.
\end{equation}
Comparing this inequality with the local realistic inequality (\ref{see-2}) we see
that the agreement between the statistical predictions of quantum mechanics and those
of our local model is both exact and complete. Thus, our local model unequivocally endorses
the statistical interpretation of the entangled singlet state, as anticipated by
Einstein \ocite{Einstein}.

\section{Derivation of the Malus's law for Sequential Spin Measurements:} 

For the above statistical interpretation of the entangled singlet state to be valid beyond
doubt, we must also show that Malus's law is respected within our local model, since  it
is respected within quantum mechanics. In this section we show that this law holds
within our local model just as exactly as it does within quantum mechanics.

Suppose we have a subensemble of spin one-half particles, prepared with a definite spin by a
polarizer, which we denote by a unit vector ${\bf p}$. Suppose next we want to calculate the
expected value of the result of a measurement of the spin component
${{\boldsymbol\sigma}\cdot{\bf a}}$, where ${\bf a}$ is a unit direction along an analyzer.
Without loss of generality, we shall choose the spins to be selected ``up'' along the
direction ${\bf p}$, with the value ${s_{\bf p}\,=\,+\,1}$. In the language of our model
this selection can be stated as ${{\boldsymbol\mu}\cdot{\bf p}\,=\,+\,I\,{\bf p}}$, which
means that the subensemble of spins prepared by means of projections of the microstates
${\boldsymbol\mu}$ along the direction ${\bf p}$ will all be spinning ``up'' in the
counterclockwise sense. Then, what we want to calculate is the expected value of finding
the spin along the direction ${\bf a}$. Since we already know the result of this
calculation from quantum mechanics, we can rephrase our question in the local realistic
terms as follows \ocite{Clauser-AJP}:
\begin{equation}
\int_{{\cal V}_3} A({\bf a},\,{\bf p},\,{\boldsymbol\mu})
\;\;d{\boldsymbol\rho}(\boldsymbol\mu)\,\stackrel{?}{=}\,
\bra{\,s_{\bf p}=+1\,}\,{\boldsymbol\sigma}\cdot{\bf a}\,\ket{\,s_{\bf p}=+1\,}\,
=\,{\bf a}\cdot{\bf p}\,, \label{2}
\end{equation}
where the RHS is the quantum mechanical expectation value, and the LHS is the average over the
microstates of a dichotomic observable ${A({\bf a},\,{\bf p},\,{\boldsymbol\mu}):=
({\boldsymbol\mu}\cdot{\bf a})\;({\boldsymbol\mu}\cdot{\bf p})}$. The latter is of course a
product of two unit bivectors, producing a quaternion of unit norm, and, as we discussed in
sections II and III, can only yield binary values ${\pm\,1}$ in any actual observation.
Thus, it can serve as a dichotomic observable for the purpose at hand. Now, since, as a physical
variable, ${A({\bf a},\,{\bf p},\,{\boldsymbol\mu})}$ has been designed to be sensitive
to what happens to the particles at the polarizer, the initial orientations of the bivectors
${{\boldsymbol\mu}\cdot{\bf a}}$ along the direction ${\bf a}$ will not be evenly
balanced between ${+\,I\,{\bf a}}$ and ${-\,I\,{\bf a}}$. Indeed, by preparation, the spins
within the initial ensemble have been filtered out to be spinning only ``up'' along the
direction ${\bf p}$, which may not be equal to the direction ${\bf a}$ in general, and hence
initially the spins will not be spinning ``up'' along the direction ${\bf a}$, unless, of
course, ${{\bf a}={\bf p}}$. In other words, the initial probability of finding the spin
``up'' along the direction ${\bf a}$ is strictly zero. Consequently, in analogy with Bell's
own derivation of the Malus's law (cf. Ref.\ocite{Bell-1964}, Eqs.${\,}$(4) to (7)), where he
preselects an appropriate subensemble of spins with definite polarization by imposing the
condition ${{\boldsymbol\lambda}\cdot{\bf p}>0}$ on his ``hidden'' observable), we must
preselect an appropriate subensemble for our scenario by assigning the weight
{\it zero} to the initial probability of orientations
${(\pm\,I\,{\bf a})\;(-\,I\,{\bf p})}$ and ${(+\,I\,{\bf a})\;(+\,I\,{\bf p})}$ of the variable
${A({\bf a},\,{\bf p},\,{\boldsymbol\mu})}$, before taking the average over the microstates
${\boldsymbol\mu}$.
This, in turn, means---equivalently---that we must assign the weight {\it one} to the initial
probability of orientation ${(-\,I\,{\bf a})\;(+\,I\,{\bf p})}$ of
the variable ${A({\bf a},\,{\bf p},\,{\boldsymbol\mu})}$. With these physical constraints in
place, the net expected value---as prescribed in equation (\ref{2}) above---can be easily
calculated as follows: 
\begin{align}
\int_{{\cal V}_3} A({\bf a},\,{\bf p},\,{\boldsymbol\mu})
\;\;d{\boldsymbol\rho}(\boldsymbol\mu)\,&=
\int_{{\cal V}_3}(-\,I\,{\bf a})\;(+\,I\,{\bf p})\;\;d{\boldsymbol\rho}(\boldsymbol\mu)\,
=\,-\,I^2\int_{{\cal V}_3} {\bf a}\,{\bf p}\;\;d{\boldsymbol\rho}(\boldsymbol\mu) \notag \\
&=\int_{{\cal V}_3} {\bf a}\,{\bf p}\;\;d{\boldsymbol\rho}(\boldsymbol\mu)\,=\,
{\bf a}\cdot{\bf p}\,+\int_{{\cal V}_3} {\boldsymbol\mu}\cdot({\bf a}\times{\bf p})
\;\;d{\boldsymbol\rho}(\boldsymbol\mu) \notag \\
&=\,{\bf a}\cdot{\bf p}\,+\,0\,, \label{what}
\end{align}
where the second and third lines follow from a use of the equations (17) to (19) of
Ref.\ocite{Christian}. Needless to say, we would arrive at the same result for the case
where the initial subensemble is preselected to be ``spin down'' along ${\bf p}$, with the
value ${s_{\bf p}\,=\,-\,1}$ and orientation ${(+\,I\,{\bf a})\;(-\,I\,{\bf p})}$.
This makes it evident that the quantum version of the Malus's law,
\begin{equation}
\bra{\,s_{\bf p}=\pm1\,}\,{\boldsymbol\sigma}\cdot{\bf a}\,\ket{\,s_{\bf p}=\pm1\,}\,
=\,{\bf a}\cdot{\bf p}\,,
\end{equation}
holds {\it exactly} within our model. Moreover, the foregoing steps from polarizer to
analyzer can be repeated at will for a sequence of measurements. For this purpose a new state 
of the system for each subsequent measurement must be prepared (cf. Ref.\ocite{Clauser-AJP}).
We shall again assume that the apparatus only lets through the ``spin up'' particles. Then, unlike
the complicated procedure that must be followed to prepare the new state in the case of Bell's
model \ocite{Clauser-AJP}, in our case all one has to do is to rotate the polarizer in the 
direction of the analyzer until the new direction of the polarizer becomes ${{\bf p'}={\bf a}}$.
In other words, after the first measurement, all one has to do is to prescribe that the apparatus
defines ${{\bf p'}={\bf a}}$ as the new polarization direction. Then the new distribution of
hidden variables will be identical to the one before the first measurement, apart from being 
rotated to the new direction ${\bf p'}$. A second measurement by an analyzer ${\bf a'}$ on a
similarly preselected subensemble ${(-\,I\,{\bf a'})\;(+\,I\,{\bf p}')}$ will then yield 
the expected value
\begin{equation}
\int_{{\cal V}_3} A({\bf a'},\,{\bf p'},\,{\boldsymbol\mu})
\;\;d{\boldsymbol\rho}(\boldsymbol\mu)\,=\,
\bra{\,s_{\bf p'}=\pm1\,}\,{\boldsymbol\sigma}\cdot{\bf a'}\,\ket{\,s_{\bf p'}=\pm1\,}
\,=\,{\bf a'}\cdot{\bf p'},
\end{equation}
again in agreement with the prediction of quantum mechanics. Thus we see that the
quantum mechanical predictions for the sequential measurements of spin-components along
arbitrary directions are duly reproduced by our model.

\section{Concluding Remarks}

Contrary to the received wisdom, Bell's theorem is not a threat to local realism. Neither is
it a curb on determinism. The counterexample constructed in Ref.\ocite{Christian} provides a
fully deterministic, common cause explanation of the EPR-Bohm correlations. In fact, it is hard
to imagine a more simple common cause than the one on which the counterexample is based---namely,
the intrinsic freedom of choice in the initial orientation of the orthogonal directions in the
Euclidean space. In the present paper we have further consolidated the conclusions of
Ref.\ocite{Christian} by demonstrating that the exact, locally causal model for the EPR-Bohm
correlations constructed therein satisfies at least eight essential requirements,
arising from either the predictions of quantum mechanics or the premises of Bell's theorem.
These requirements, as listed in the Introduction, include the locality condition of Bell, and
hence by respecting them our model fully endorses the view that the quantum mechanical description
of reality is incomplete \ocite{EPR}.
Moreover, since this view is reinforced by three different local realistic derivations of the
violations of the CHSH inequality \ocite{Christian}\ocite{reply}, and since all three of them
agree with the corresponding predictions of quantum mechanics in {\it quantitatively} precise
manner, the statistical interpretation of the entangled singlet state becomes the most natural
interpretation of this state, as anticipated by Einstein. It is therefore hoped
that---strengthened by the results of the present paper---the counterexample of
Ref.\ocite{Christian} would rejuvenate the search for a unified, locally causal basis for
the whole of physics, as envisaged by Einstein \ocite{Einstein}.

\acknowledgments

This paper has been inspired and shaped by the largely skeptical questions and comments
concerning
Ref.\ocite{Christian} from many friends and colleagues, including Paul Busch, Marek Czachor,
Laurent Freidel, Richard Gill, Lucien Hardy, Marc Holman, Matthew Leifer, Owen Maroney,
Peter Morgan, Marcin Pawlowski, Markus Penz, Kevin Resch, Michael Seevinck, Abner Shimony,
Lee Smolin, Rafael Sorkin, Ward Struyve, Chris Timpson, Gregor Weihs, and Hans Westman. The
research for this paper was carried out during a visiting appointment at the Perimeter
Institute for Theoretical Physics, Canada, whose generous hospitality and support are
gratefully acknowledged.

\end{document}